\documentclass[a4paper,12pt]{article}

\usepackage[utf8]{inputenc}
\usepackage{amsmath,amssymb,amsfonts}
\usepackage{dsfont}

\usepackage{tikz}
\usetikzlibrary{matrix}

\usepackage[colorlinks,urlcolor  = black]{hyperref}
\usepackage{ntheorem}
\usepackage[capitalize,nameinlink]{cleveref}

\usepackage{enumerate}

\usepackage{mathrsfs}
\usepackage{ulem}
\usepackage[makeroom]{cancel}
\usepackage{graphicx}
\usepackage{caption}
\usepackage{amsopn} 

\usepackage{pdfpages}  

\hypersetup{linkcolor={black},citecolor=black}

\numberwithin{figure}{section}

\renewcommand{\epsilon}{\varepsilon}
\renewcommand{\theta}{\vartheta}
\renewcommand{\kappa}{\varkappa}
\renewcommand{\rho}{\varrho} 
\renewcommand{\phi}{\varphi}

\newcommand{\comm}[1]{}

\crefname{deff}{Definition}{Definitions}

\crefname{rem}{Remark}{Remarks}

\crefname{lem}{Lemma}{Lemmas}
\newtheorem{theo}{Theorem}
\crefname{theo}{Theorem}{Theorems}
\newtheorem{cor}{Corollary}
\crefname{cor}{Corollary}{Corollaries}

\crefname{ex}{Example}{Examples}

\crefname{prop}{Proposition}{Propositions}
\crefname{figure}{Figure}{Figures}

\crefname{equation}{}{}

\newcommand{\pd}[2]{\ifthenelse{\equal{\detokenize{#1}}{\detokenize{•}}}
						{\frac{\partial}{\partial #2}}
						{\frac{\partial #1}{\partial #2}}
					}
\newcommand{\Sum}[3]{\ifthenelse{\equal{\detokenize{#1}}{\detokenize{•}}}
							{\sum_{#2}{#3}}  
							{\sum^{#1}_{#2}{#3}}	
					}

\newcommand{\rom}[1]{%
  \textup{\uppercase\expandafter{\romannumeral#1}}%
}

\usepackage{setspace}
\begin{document}

\title{\protect\vspace*{-2cm}Complete Classification of Local Conservation Laws for Generalized Cahn--Hilliard--Kuramoto--Sivashinsky Equation}

	\author{Pavel Holba\\[2mm]
		Mathematical Institute, Silesian University in Opava,\\
                  Na Rybn\'\i{}\v{c}ku 1, 74601 Opava, Czech Republic\\
	E-mail: {\tt pavel.holba@math.slu.cz}}

\maketitle

\begin{abstract}\setstretch{1.0}
In the present paper we consider nonlinear multidimensional Cahn--Hilliard and Kuramoto--Sivashinsky equations that have many important applications in physics and chemistry, and a certain natural generalization of these equations.

For an arbitrary number of spatial independent variables we present a complete list of cases when the generalized Cahn--Hilliard--Kuramoto--Sivashinsky equation admits nontrivial local conservation laws of any order, and for each of those cases we give an explicit form of all the local conservation laws of all orders modulo trivial ones admitted by the equation under study. 

In particular, we show that the original Kuramoto--Sivashinsky equation admits no nontrivial local conservation laws, and find all nontrivial local conservation laws for the Cahn--Hilliard equation.\looseness=-1

\medskip

\noindent{\bf Keywords}: {conservation laws; Cahn--Hilliard equation; Kuramoto--Sivashinsky equation}
\end{abstract}  	
	


	\section*{Introduction}
	Below we study the generalized Cahn--Hilliard--Kuramoto--Sivashinsky equation, a PDE in $n+1$ independent variables $t,x_1\ldots,x_n$ and one dependent variable $u$ of the form
	\begin{equation}
		u_t= a \Delta^2u+b(u)\Delta u+f(u)|\nabla u|^2+g(u),
		\label{Holba:GKS}
	\end{equation}
	where $a$ is a nonzero constant, $b,f,g$ are smooth functions of $u$, $\Delta=\sum_{i=1}^n\partial^2/\partial x_i^2$ is the Laplace operator, $|\nabla u|^2=\sum_{i=1}^n(\partial u/\partial x_i)^2$, and $n$ is an arbitrary natural number. 

	
	Equation \cref{Holba:GKS} is a natural generalization of two well-known equations, the Cahn--Hilliard equation
\begin{equation}
		u_t=c_1\Delta(u^3-u+c_2\Delta u),
		\label{Holba:CH}
	\end{equation}
where $c_1,c_2$ are constants (this equation can be obtained from \cref{Holba:GKS} by setting $a=c_1c_2$, $b=c_1(3u^2-1)$, $f=6c_1u$ and $g=0$), and the Kuramoto--Sivashinsky equation 
\begin{equation}
		u_t + \Delta^2 u + \Delta u + |\nabla u|^2/2=0,
		\label{Holba:KS}
	\end{equation}
obtained from \cref{Holba:GKS} by setting $a=-1$, $b=-1$, $f=-1/2$ and $g=0$. 

Equations \cref{Holba:CH} and  \cref{Holba:KS} arise in a variety of physical, chemical and biological contexts, with the Cahn--Hilliard equation \cref{Holba:CH} describing inter alia the process of phase separation in a binary alloy and having applications in many areas such as complex fluids, interfacial fluid flow, polymer science, spinodal decomposition and tumor growth simulation, see e.g \cite{JSYSD,BDS,LS} and references therein, and the Kuramoto--Sivashinsky equation \cref{Holba:KS} describing inter alia flame propagation, reaction-diffusion systems and unstable drift waves in plasmas, see e.g.\ \cite{GF, HN} and references therein. Let us also note that equation \cref{Holba:KS} exhibits chaotic behavior and is an important model in the study of chaotic phenomena, cf.\ e.g.\ \cite{HN, PSm}.\looseness=-1 

	

While many authors  have considered various generalizations of equations \cref{Holba:KS,Holba:CH}, see e.g.\cite{CMZ,KS} for equation \cref{Holba:CH} and \cite{BD, NK, Rosa} for equation \cref{Holba:KS}, there is no generally accepted definition of the term `generalized' in relation to these equations. 

Our goal is to present a complete description of local conservation laws for \cref{Holba:GKS} for all cases when they exist. To the best of our knowledge, this was not yet done, especially for an arbitrary number $n$ of the space variables, although e.g.\ some partial results on conservation laws of several different generalizations of the Kuramoto--Sivashinsky equation for $n=1,2$ are known, see e.g.\ \cite{BD, GNO, Rosa}. 

Recall that conservation laws have numerous important applications, cf.\ e.g.\ \cite{BP, CS, FT, Olver,S, T}, including for instance improving numerical solving the PDE under study using discretizations respecting known conservation laws, see e.g.\ \cite{BM,FGH} and references therein, and construction of symmetries using Noether theorem and other methods, see for example \cite{KVV, MS, Olver, S} and references therein. 

On the other hand, obtaining a complete description of local conservation laws for a given PDE, rather than finding a few low-order ones, is a difficult task that was achieved only for a rather small number of examples, cf.\ e.g.\ \cite{H,  i, OP, SV, JV} and references therein. 

The rest of the paper is organized as follows. In Section~\ref{pre} we recall some necessary definitions and set up notation, in Section~\ref{mai} the main results are presented, while their proofs are given in Section~\ref{pro}. 
	


\section{Preliminaries}\label{pre}
In this section we just recall standard definitions and introduce some notation mostly following \cite{Olver} but restricted to the case of the equation under study, i.e., \cref{Holba:GKS}.

First of all, we define a \textit{differential function} as a smooth function depending on $t,x_1\ldots,x_n,u$ and finitely many derivatives of $u$.

The symbols $D_t$ and $D_{x_i}$ will stand below for the so-called total derivatives, see e.g.\ \cite{i, KVV, Olver} for more details on those.

	A \textit{local conservation law} for \cref{Holba:GKS} is a differential expression 
	\begin{equation}
		D_t(T)+\textnormal{Div}X
		\label{ZZ}
	\end{equation}
	which vanishes on all smooth solutions of \cref{Holba:GKS}. Here $X=(X_1,\ldots,X_n)$, $\textnormal{Div}X=\sum_{i=1}^{n}{D_{x_i}(X_i)}$ stands for total divergence and $T,X_1,\ldots,X_n$ are differential functions. 

The function $T$ in \cref{ZZ} is called a \textit{density} and the vector function $X$ is called a \textit{flux} for the conservation law under study.
	
A local conservation law is said to be \textit{trivial}  if either $T$ and $X$ themselves vanish on all (smooth) solutions of \cref{Holba:GKS} or $D_t(T)+\textnormal{Div}X \equiv 0$ holds  identically no matter whether \cref{Holba:GKS} holds.
	
Two local conservation laws are said to be \textit{equivalent} if their difference is a trivial conservation law.
	
A local conservation law \cref{ZZ} for  \cref{Holba:GKS} is in \textit{characteristic form} if
	$$
		D_t(T)+\textnormal{Div}X=Q\cdot (u_t-( a \Delta^2u+b(u)\Delta u+f(u)|\nabla u|^2+g(u))),
	$$
	where the differential function $Q$ is called a \textit{characteristic} of the conservation law in question. 
	
Note that a local conservation law for \cref{Holba:GKS} is nontrivial if and only if its characteristic is nonzero. 
		
Below we shall deal only with local conservation laws, so, whenever a conservation law is mentioned, we mean a local conservation law unless explicitly stated otherwise.

In what follows we assume without loss of generality, cf.\ e.g.\ \cite{Olver}, that  
$X_i$ and $T$ do not depend on the $t$-derivatives of $u$ or mixed derivatives of $u$ involving both $x_i$ and $t$, as disallowing such dependence only removes certain trivial parts from the conservation laws under study because of evolutionary nature of \cref{Holba:GKS}.\looseness=-1 

\section{Main results}\label{mai}

\begin{theo}\label{NonexistenceThm}
	Let equation \cref{Holba:GKS} with $a\neq 0$ and smooth functions $b(u),f(u)$, and $g(u)$ satisfy one of the following sets of conditions:
	\begin{enumerate}
		\item $\displaystyle f \neq \frac{\partial b}{\partial u}$,
		\item $\displaystyle f = 0$, $\displaystyle \frac{\partial^2 g}{\partial u^2} \neq 0$ and $b$ is constant,
		\item $\displaystyle f = \frac{\partial b}{\partial u} \neq 0$, $\displaystyle \frac{\partial^2 g}{\partial u^2}\neq 0$ and	$\displaystyle \frac{\partial^3 g}{\partial u^3}\frac{\partial b}{\partial u}-\frac{\partial^2 g}{\partial u^2}\frac{\partial^2 b}{\partial u^2} \neq 0$.
	\end{enumerate}

	Then equation \cref{Holba:GKS} admits no nontrivial local conservation laws.
\end{theo}

	In particular,  we immediately get the following important corollary of the above theorem:
	\begin{cor}
		The Kuramoto--Sivashinsky equation \cref{Holba:KS} has no nontrivial local conservation laws.
	\end{cor}

Thus, under the conditions of Theorem \ref{NonexistenceThm} equation \cref{Holba:GKS} and, in particular, the original Kuramoto--Sivashinsky equation \cref{Holba:KS}, admit no nontrivial local conservation laws. This result has a number of important consequences. For one, as there are no local conservation laws in the cases under study, while discretizing \cref{Holba:KS,Holba:GKS} under the assumptions of Theorem~\ref{NonexistenceThm} in order to solve the equations in question numerically, one does not have to worry about consistency of the chosen discretization with the conservation laws, cf.\ the discussion in the introduction and in \cite{H} and \cite{FRK}.
	\begin{cor}
		The generalized Cahn--Hilliard equation (cf.\ eq.\  (1.7)) in \cite{CMZ}),
		\begin{equation}
			u_t+\Delta^2u-\Delta\tilde{f}(u)+\tilde{g}(u)=0 
			\label{GCH}
		\end{equation}
		obtained from \cref{Holba:GKS} by setting $a=-1,\, b=\displaystyle\frac{\partial \tilde{f}}{\partial u},\, f=\displaystyle\frac{\partial^2 \tilde{f}}{\partial u^2},\, g=-\tilde{g}$ admits no nontrivial local conservation laws provided one of the following sets of conditions holds:
		\begin{enumerate}
			\item $\displaystyle \frac{\partial^2 \tilde{f}}{\partial u^2} = 0$ and $\displaystyle\frac{\partial^2 \tilde{g}}{\partial u^2} \neq 0$
			\item $\displaystyle \frac{\partial^2 \tilde{f}}{\partial u^2} \neq 0$, $\displaystyle\frac{\partial^2 \tilde{g}}{\partial u^2} \neq 0$ and $\displaystyle \frac{\partial^3 \tilde{g}}{\partial u^3}\frac{\partial^2 \tilde{f}}{\partial u^2}-\frac{\partial^2 \tilde{g}}{\partial u^2}\frac{\partial^3 \tilde{f}}{\partial u^3} \neq 0$.
		\end{enumerate}				
		 
	\end{cor}


\begin{theo}\label{clt}
If $\displaystyle f = \frac{\partial b}{\partial u}$  then \cref{Holba:GKS} with $a\neq 0$ and smooth functions $b(u),f(u)$, and $g(u)$ admits nontrivial local conservation laws, all of which are listed below modulo trivial ones, only in the following three cases:

	I. Let $f = 0$, $b=c_1$ and $g=c_2 u+c_3$,where $c_1, c_2, c_3$ are arbitrary constants. Then the densities of all nontrivial local conservation laws for \cref{Holba:GKS} have the form
	\begin{equation}
		T^\rom{1}=uQ^\rom{1}+\frac{1}{c_2}Q^\rom{1}	
	\end{equation}
where $Q^\rom{1}=Q^\rom{1}(t,x_1,\dots,x_n)$ is any (smooth) solution of the linear PDE with constant coefficients 
	\begin{equation}\label{linGKS}
		\frac{\partial Q^\rom{1}}{\partial t}+a\Delta^2(Q^\rom{1})+c_2Q^\rom{1}+c_1\Delta(Q^\rom{1})=0.
\end{equation}

	II. Let $\displaystyle f =\frac{\partial b}{\partial u} \neq 0$ and $g=c_2u+c_3$, where $c_2$ and $c_3$ are arbitrary constants. Then the densities of all nontrivial  
 local conservation laws for \cref{Holba:GKS} have the form\vspace{-2mm}
\begin{equation}
		T^\rom{2}=\tilde{Q}^\rom{2}e^{-c_2t}u\vspace{-2mm}
	\end{equation}
	where $\tilde{Q}^\rom{2}=\tilde{Q}^\rom{2}(x_1,\ldots,x_n)$  is any (smooth) solution of the linear Laplace equation 
\begin{equation}\label{lap}\Delta\tilde{Q}^\rom{2}=0.\end{equation}

	III. Let $\displaystyle f =\frac{\partial b}{\partial u} \neq 0$, $\displaystyle \frac{\partial^2 g}{\partial u^2}\neq 0$ and $\displaystyle\frac{\partial^3 g}{\partial u^3}\frac{\partial b}{\partial u}=\frac{\partial^2 g}{\partial u^2}\frac{\partial^2 b}{\partial u^2}.$
Then there exist constants $c_4 \neq 0$ and $c_5$ such that $\displaystyle b = \frac{1}{c_4}\frac{\partial g}{\partial u}+c_5$, and the densities of all nontrivial  
 local conservation laws for \cref{Holba:GKS} have the form
\begin{equation}
		T^\rom{3}=e^{(c_4c_5-ac_4^2)t}u\tilde{Q}^\rom{3}	
	\end{equation}	
	where $\tilde{Q}^\rom{3}=\tilde{Q}^\rom{3}(x_1,\ldots,x_n)$ is any (smooth) solution of a linear PDE with constant coefficients 
\begin{equation}\label{modlap}\Delta\tilde{Q}^\rom{3}+c_4\tilde{Q}^\rom{3}=0.\end{equation}
\end{theo}

The following result is readily verified by straightforward computation.

\begin{cor}\label{fluxes} The fluxes for the conservation laws listed in the above theorem have the following form, up to the obvious trivial contributions:
  
I.  If $f = 0$, $b=c_1$ and $g=c_2u+c_3$, where $c_1, c_2, c_3$ are constants, then the  fluxes associated with the conservation laws with densities of the form $T^\rom{1}$ 
are
	\begin{equation}
	\begin{aligned}
		X_i^\rom{1}= & c_1\left(\left[\frac{1}{c_2}+u\right]\frac{\partial Q^\rom{1}}{\partial x_i}-u_{x_i}Q^\rom{1}\right)+\\
		& +a\sum_{j=1}^n{\left[\left(\frac{1}{c_2}+u\right)\frac{\partial^3 Q^\rom{1}}{\partial x_i \partial x_j^2}-u_{x_i}\frac{\partial^2 Q^\rom{1}}{\partial x_j^2}+u_{x_jx_j}\frac{\partial Q^\rom{1}}{\partial x_i}-u_{x_jx_jx_i}Q^\rom{1}\right]}.
		\end{aligned}
	\end{equation}
	II. If $f = \displaystyle\frac{\partial b}{\partial u} \neq 0$ and $g=c_2u+c_3$, where $c_2,c_3$ are constants, then the  fluxes associated with the conservation laws with densities of the form $T^\rom{2}$  are
	\begin{equation}
	\begin{aligned}
		X_i^\rom{2}= & e^{-c_2t}\Bigg[\sum_{j=1}^n{\left(a\frac{\partial\tilde{Q}^\rom{2}}{\partial x_i}u_{x_jx_j}-au_{x_jx_jx_i}\tilde{Q}^\rom{2}+\tilde{Q}^\rom{2}bu_{x_j}+\frac{\partial\tilde{Q}^\rom{2}}{\partial x_j}\tilde{b}\right)}-\\
		& - c_3x_i\tilde{Q}^\rom{2}-\frac{c_3}{2}x_i^2\frac{\partial\tilde{Q}^\rom{2}}{\partial x_i}\Bigg],
	\end{aligned}
	\end{equation}
	where $\tilde{b}= \tilde{b}(u)$ is such that $\partial \tilde{b}/\partial u=b$.

	III. If  $\displaystyle f =\frac{\partial b}{\partial u} \neq 0$, $\displaystyle \frac{\partial^2 g}{\partial u^2}\neq 0$ and $\displaystyle\frac{\partial^3 g}{\partial u^3}\frac{\partial b}{\partial u}=\frac{\partial^2 g}{\partial u^2}\frac{\partial^2 b}{\partial u^2},$ so
$\displaystyle b = \frac{1}{c_4}\frac{\partial g}{\partial u}+c_5$ for suitable constants $c_4$ and $c_5$, with $c_4\neq 0$, then the fluxes associated with the conservation laws with densities of the form $T^\rom{3}$  are
	\begin{equation}
	\begin{aligned}
		X_i^\rom{3}	 = & e^{(c_4c_5-ac_4^2)t}\Bigg(a\tilde{Q}^\rom{3}u_{x_i}-a\frac{\partial\tilde{Q}^\rom{3}}{\partial x_i}u- c_5\tilde{Q}^\rom{3}u_{x_i}+c_5u\frac{\partial \tilde{Q}^\rom{3}}{\partial x_i}-\frac{1}{c_4}\tilde{Q}^\rom{3}\frac{\partial g}{\partial u}u_i + \\[5mm]
		& + \frac{1}{c_4}g\frac{\partial\tilde{Q}^\rom{3}}{\partial x_i} + a\sum_{j=1}^n{\left[\frac{\partial \tilde{Q}^\rom{3}}{\partial x_i}u_{x_jx_j}-\tilde{Q}^\rom{3}u_{x_ix_jx_j}\right]}  \Bigg).
	\end{aligned}
	\end{equation}
\end{cor}

\begin{cor}
	The only nontrivial local conservation laws for the Cahn--Hilliard equation \cref{Holba:CH}
		$$
		u_t=c_1\Delta(u^3-u+c_2\Delta u)
		$$ 
        with $c_1\neq 0$ and $c_2\neq 0$ are, modulo trivial ones, those with the densities of the form $Q(x_1,\dots,x_n)u$ where $Q$ satisfies the Laplace equation $\Delta Q=0$
\end{cor}

	Local conservation laws of generalized Cahn--Hilliard equation \cref{GCH} whenever they exist can be obtained by substituting 
	$$a=-1,\, b=\displaystyle\frac{\partial \tilde{f}}{\partial u},\, f=\displaystyle\frac{\partial^2 \tilde{f}}{\partial u^2},\, g=-\tilde{g}$$
	into general results from \cref{clt} and \cref{fluxes}.


\section{Proof of the main results}\label{pro}


Since the proofs of both  Theorems \ref{NonexistenceThm} and \ref{clt} are based on the analysis of the same determining equation for the characteristics of conservation laws for \cref{Holba:GKS} we will prove the theorems in question simultaneously. 

{\noindent{\textit{Proof of Theorems \ref{NonexistenceThm} and \ref{clt}.}} Let a differential function $Q$ be a characteristic of a local conservation law for \cref{Holba:GKS}. Just like for $X_i$ and $T$, cf.\ Section~\ref{pre}, we shall assume without loss of generality that $Q$ depends only on $t,x_i,u$ and finitely many $x$-derivatives of $u$ but do not depend on $t$-derivatives of $u$ and derivatives of $u$ involving both $t$ and $x_i$, cf.\ e.g.\ \cite[Ch.\ 4]{Olver}. 
	
Then for $Q$ to be a characteristic of a local conservation law for \cref{Holba:GKS} it is necessary that\looseness=-1 
	\begin{equation}
	\begin{aligned}
		& D_t(Q)+a\sum_{i,j=1}^n{D_{x_ix_ix_jx_j}(Q)}+\left(2\frac{\partial b}{\partial u}-2f\right)\sum_{i=1}^n{\Big[u_{x_ix_i}Q+u_{x_i}D_{x_i}(Q)\Big]}+\\
		& +\left(\frac{\partial^2 b}{\partial u^2}-\frac{\partial f}{\partial u}\right)\sum_{i=1}^n{u_{x_i}^2Q}+\frac{\partial g}{\partial u}Q+b\sum_{i=1}^n{D_{x_ix_i}(Q)}=0.
	\end{aligned}
	\label{cond_prvni}
	\end{equation}

Equation \cref{Holba:GKS} satisfies the conditions of Theorem 6 from \cite{i} with $N=-1$ and hence $Q$ can depend at most on $t,x_1\ldots,x_n$. 

It is easily seen that, without loss of generality we can assume that the density $T$ of the associated conservation law depends at most on $t,x_1\ldots,x_n$ and $u$ and we have (cf. e.g. \cite[ p. 349]{Olver})
\begin{equation}\label{tq}
\partial T/\partial u=Q,
\end{equation}
whence we can readily find $T$ if given $Q$. 

\newpage 
 
With this in mind it is readily verified 
that $Q$ is a characteristic of conservation law for \cref{Holba:GKS} if and only if it satisfies the following simplification of condition \cref{cond_prvni}:
	\begin{equation}
	\begin{aligned}
		& \frac{\partial Q}{\partial t}+a\sum_{i, j=1}^n{\frac{\partial^4 Q}{\partial x_i^2 \partial x_j^2}+\left(2\frac{\partial b}{\partial u}-2f\right)\sum_{i=1}^n{\Big(u_{x_i x_i}Q+u_{x_i}\frac{\partial Q}{\partial x_i}}\Big)}+\\
		& +\left(\frac{\partial^2 b}{\partial u^2}-\frac{\partial f}{\partial u}\right)\sum_{i=1}^n{u_{x_i}^2Q}+\frac{\partial g}{\partial u}Q+b\sum_{i=1}^n{\frac{\partial^2 Q}{\partial x_i^2}}=0.
	\end{aligned}
	\label{necessary_cond}
	\end{equation}
	
As $Q$ is independent of $u$ and its derivatives, applying $\partial/\partial u_{x_ix_i}$ to \cref{necessary_cond} for any $i$ we get 
	\begin{equation}
		\left( 2\frac{\partial b}{\partial u} - 2f \right)Q=0,
	\end{equation}
	which means that we must have
	\begin{equation}
		f=\frac{\partial b}{\partial u}
		\label{cond1}
	\end{equation}	
	otherwise there exist no nontrivial conservation laws. This establishes part 1 of Theorem \ref{NonexistenceThm}.
	
For the rest of the proof we assume that \cref{cond1} holds.

 Using \cref{cond1} one can simplify equation \cref{necessary_cond} to
	\begin{equation}
		\frac{\partial Q}{\partial t}+a\sum_{i,j=1}^n{\frac{\partial^4 Q}{\partial x_i^2 \partial x_j^2}}+\frac{\partial g}{\partial u}Q+b\sum_{i=1}^n{\frac{\partial^2 Q}{\partial x_i^2}}=0.
		\label{necessary_cond2}
	\end{equation}	
	Upon applying $\partial/\partial u$ to \cref{necessary_cond2} we get
	\begin{equation}
		\label{cond2}
		\frac{\partial^2 g}{\partial u^2}Q+\frac{\partial b}{\partial u}\sum_{i=1}^n{\frac{\partial^2 Q}{\partial x_i^2}}=0.
	\end{equation}
	We can split the analysis of \cref{cond2} into three cases labelled as A, B and C.

	{\sl Case A:} $\displaystyle\frac{\partial b}{\partial u}=0$, meaning that $b$ is constant and $f=0$.
	
	\smallskip

    Then from \cref{cond2} we get
	\begin{equation}
		\frac{\partial^2 g}{\partial u^2}Q=0,
		\label{cond3}
	\end{equation}
	which means that we must have
	\begin{equation}
		\frac{\partial^2 g}{\partial u^2}=0
	\end{equation}
	otherwise there exist no nontrivial conservation laws. This establishes part 2 from Theorem \ref{NonexistenceThm}. 

Setting $b=c_1$ and $g=c_2u+c_3$, where $c_1, c_2 $ and $c_3$ are arbitrary constants, we can simplify \cref{necessary_cond2} to 
	\begin{equation}
		\frac{\partial Q}{\partial t}+a\sum_{i,j=1}^n{\frac{\partial^4 Q}{\partial x_i^2 \partial x_j^2}}+c_2Q+c_1\sum_{i=1}^n{\frac{\partial^2 Q}{\partial x_i^2}}=0.
	\end{equation}

\newpage

	Now using \cref{tq} we can readily find the asociated $T$, and thus establish part I from Theorem \ref{clt}. For convenience we denote in Theorem \ref{clt} the relevant $Q$ and $T$ as $Q^{\rom{1}}$ and $T^{\rom{1}}$ to indicate their relation to part I of the theorem in question, and adopt similar notation for parts II and III. Note that the way we used the residual freedom in the choice of the form of $T^{\rom{1}}$, making it inhomogeneous in $u$, is motivated by the desire to keep the form of the associated flux components $X^{\rom{1}}_i$ reasonably simple.
	
	\medskip  

	{\sl Case B:} $\displaystyle\frac{\partial b}{\partial u}\neq 0$ but $\displaystyle\frac{\partial^2 g}{\partial u^2}Q=0$. 

\smallskip

Then from \cref{cond2} we get
	\begin{equation}
		\sum_{i=1}^n{\frac{\partial^2 Q}{\partial x_i^2}}=0.
		\label{cond4}
	\end{equation}
	Setting $g=c_2u+c_3$ and using equation \cref{cond4} we can simplify \cref{necessary_cond2} to
	\begin{equation}
		\frac{\partial Q}{\partial t}+c_2Q=0.
	\end{equation}
	Solving these we get $Q=e^{-c_2t}\tilde{Q}$, where $\tilde{Q}$ is function of independent variables $x_1,\ldots,x_n$ which needs to satisfy the Laplace equation $\Delta(\tilde{Q})=0$, establishing part II from Theorem \ref{clt} upon another application of \cref{tq}.


	{\sl Case C:} $\displaystyle\frac{\partial b}{\partial u}\neq 0$ and $\displaystyle\frac{\partial^2 g}{\partial u^2}\neq 0$. 
	
\smallskip	

    Then we can divide \cref{cond2} by $\partial b/\partial u$ and apply $\partial/\partial u$, whence we get
	\begin{equation}
	\label{cond5}
		\left(\displaystyle\frac{\partial^3 g}{\partial u^3}\frac{\partial b}{\partial u}-\frac{\partial^2 g}{\partial u^2}\frac{\partial^2 b}{\partial u^2}\right)Q\left/{\left(\displaystyle\frac{\partial b}{\partial u}\right)^2}\right.=0,
	\end{equation}
	which means that we must have
	\begin{equation}
	\label{cond6}
	\frac{\partial^3 g}{\partial u^3}\frac{\partial b}{\partial u}=\frac{\partial^2 g}{\partial u^2}\frac{\partial^2 b}{\partial u^2}
	\end{equation}
	otherwise there exist no nontrivial conservation laws. This establishes part 3 from Theorem \ref{NonexistenceThm}. 

	Assuming \cref{cond6} holds, we can solve \cref{cond6} for $b$: 
	\begin{equation}
	b=\frac{1}{c_4}\frac{\partial g}{\partial u}+c_5,
	\label{cond7}
	\end{equation}
where $c_4\neq 0$ and $c_5$ are arbitrary constants,
	and simplify \cref{cond2} to 
	\begin{equation}
		Q+\frac{1}{c_4}\sum_{i=1}^n{\frac{\partial^2 Q}{\partial x_i^2}}=0.
		\label{cond8}
	\end{equation}
	Using \cref{cond7} and \cref{cond8} we can simplify \cref{necessary_cond2} to
	\begin{equation}
		\frac{\partial Q}{\partial t}+(a c_4^2-c_4 c_5)Q=0.
	\end{equation}
      Hence $Q=e^{-(ac_4^2-c_4c_5)t}\tilde{Q}$, where $\tilde{Q}$ is a (smooth)  function of independent variables $x_1,\ldots,x_n$ which needs to satisfy $c_4 \tilde{Q}+\Delta\tilde{Q}=0$. Making use of \cref{tq} yields the associated density $T$, thus  establishing part III of Theorem~\ref{clt} and completing the proof. $\Box$\\[1mm]

Note that while for $n>1$ in all three cases listed in Theorem~\ref{clt} equation \cref{Holba:GKS} admits infinitely many local conservation laws, for $n=1$ the situation is strikingly different: in the case I there is still infinitely many nontrivial local conservation laws, while in the cases II and III there are just two, because for $n=1$ equations \cref{lap} and \cref{modlap} become linear {\sl ordinary} differential equations of second order while \cref{linGKS} remains a  linear {\sl partial} differential equation in two independent variables. Notice also that for $n=1$ the results of case I of Theorem~\ref{clt} readily follow, up to a shift of $u$ by a suitable constant to turn the special case of \cref{Holba:GKS} under study into a linear homogeneous PDE, from Theorem 3 of \cite{PS}.\looseness=-1 

In fact, the presence of infinitely many local conservation laws in case I of Theorem~\ref{clt} is not unexpected, because this is a degenerate case of sorts, when \cref{Holba:GKS} becomes a {\sl linear inhomogeneous} partial differential equation. Even for $n>1$ there still is a significant difference between the cases in Theorem \ref{clt} as it is readily verified that for case I the associated infinite set of local conservation laws is parametrized by an arbitrary (smooth) function of $n$ independent variables $x_1,\dots,x_n$ while for cases II and III the associated infinite sets of associated local conservation laws are parameterized by pairs of arbitrary (smooth) functions of $n-1$ variables.  	
\subsection*{Acknowledgments}	
	This research was supported in part by the Specific Research Grant SGS/13/2020 of Silesian University in Opava.\looseness=-1

	I would like to thank Artur Sergyeyev for the patient guidance, encouragement and advice. 
\addcontentsline{toc}{section}{\protect\numberline{}References}%

\end{document}